# How do you Command an Army of Intelligent Things?

Alexander Kott, US Army Research Laboratory, and David S. Alberts, Institute for Defense Analysis

Within a decade, probably less, we will need to find ways to work effectively with ever growing numbers of intelligent things, including robots and intelligent agents. We will have to, because we will encounter them with increased frequency in our homes and places of work.  The rapid development of information and communications technologies – such as networks of intelligent things -- that has occurred in recent decades has been transformational.  A look at the changing composition of the Fortune 100 over the past century should be sufficient to see how those that have leveraged the power of information and automation have fared.   Despite the emergence of networked enterprises, one could argue that this shift, however profound, has not fundamentally affected the nature of our organizations, especially the management processes of our organizations.

This is about to change.

The source of this disruptive development is the advent and rapid adoption of advanced forms of 'machine intelligence,' in the form of intelligent things, robots and software agents, that are capable of, and will possess, a degree of autonomy.   The networked workforce of the near future will thus consist of not only interconnected and interdependent humans but also of intelligent things.   The humans among them will find themselves to be merely a particular specie of intelligent entities, in fewer and fewer numbers in relation to other intelligent things.  And, at least some of these intelligent specie need to be considered, from a management perspective, as entities with decision-making responsibilities, similar to human individuals to be accounted for in the design of our organizations.

This raises a number of challenging issues, none more compelling and urgent than finding an answer to the question "How to manage this new organizational form?"   It is hard enough to effectively command or manage a purely human organization.  It is even harder to do so with the agility required for the complex and dynamic environments in which we must operate. How can we rise to the occasion?

Let us consider these issues in a particularly challenging domain of human endeavor – warfare [4]. Command and Control (C2) is the term applied to management or governance of military organizations and endeavors. The question then is "How can we command and control an army of intelligent things? Humans and other intelligent entities each can bring to the table different strengths in efforts to accomplish key C2 or management functions. We will need to understand and leverage the comparative strengths and weaknesses of human cognition and machine intelligence in order to design an approach to C2 that is appropriate for this new organizational form.

Figure 1 depicts the essential C2 functions (command, control, sensemaking, execution, situation monitoring) necessary to create desired effects in a dynamic context [1]. Successfully accomplishing these functions requires the ability to collect, process, and share information. We will consider how human and other intelligent entities can best contribute to ensuring that the decision makers, whether human or machine, have the information they require and make good use of this information to accomplish C2 functions.

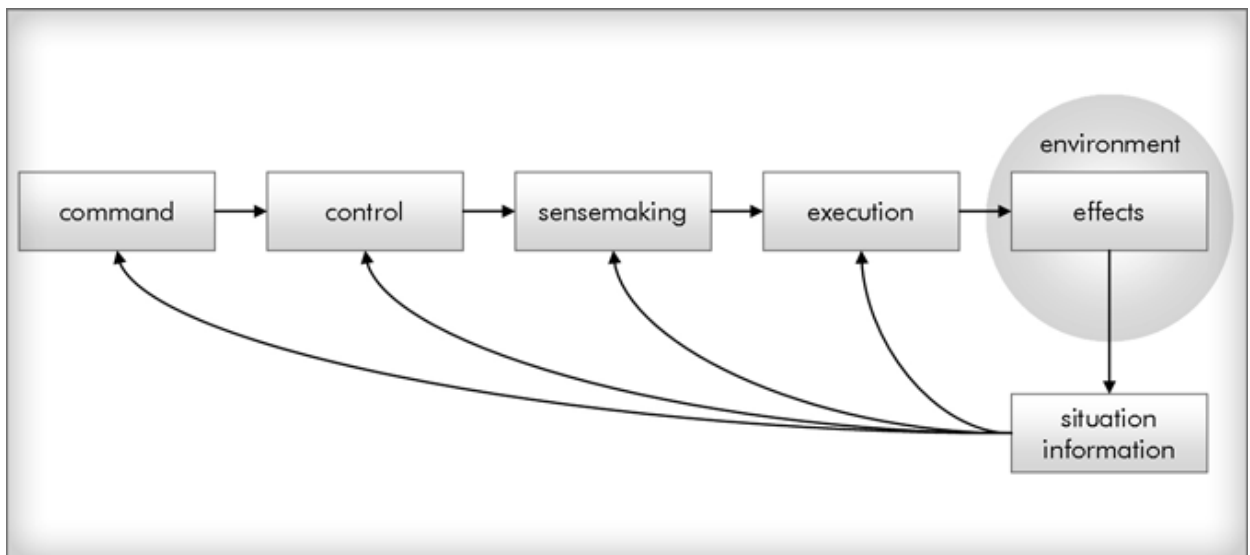

*Figure 1. To create desired effects in a complex, dynamic environment, an organization must execute command and control functions: command, control, sensemaking, execution, and situation monitoring.*

## Command

The function of command includes establishing intent and creating the conditions that enable the organization to achieve the desired results. Intent consists setting goals and priorities, establishing rules, and setting constraints. Creating the conditions for success include assigning roles and responsibilities and defining relationships.

In establishing and expressing intent, humans exhibit their share of shortcomings. Humans tend to be vague and unspecific about the intent. They also tend to make numerous assumptions, more often implicit and unspoken than explicitly articulated, and may engage in self-serving interpretations regarding the priorities. Personal agendas and group-think often dominate. Unlike humans, intelligent things are far more likely to stick to well defined, formally stated goals and priorities, with no personal agendas. Inconsistent goals will be identified and highlighted. Humans, however, will find it difficult and infuriating to provide the level of formality and specificity in matters of intent, goals and priorities that intelligent things require.

Every decision-making organization strives to determine roles, responsibilities and relationships within the organization, in order to better match the requirements of the problems and solutions at hand.   Recent research has assembled the empirical evidence to support the thesis that there is not one size fits all approach to Command and Control that is appropriate for all missions and circumstances and that military organizations need to be prepared with options, different ways allocating decision rights, different patterns of interact, and access to information. [8]

In a less formal and more ad hoc manner, humans within a decision-making network continually have been modifying their own roles, tasks and relations to other nodes of the networks, formally or informally. They also attempt to modify those of others, either by ordering organizational changes to the subordinates, or trying to influence other nodes by a variety of means. Humans also try to manage trust and cooperative relations with other nodes. Informal tasks forces or ad hoc groups are created, new experts or champions are discovered, etc. In many situations, this results in adjustments to the structure of organization that exhibits a better match to the problem at hand. In other, less satisfactory cases, the organization drives itself into an unproductive mode. How well would intelligent things fair in these activities?  We suspect the current generation of intelligent things would fair rather poorly.

In the foreseeable future, intelligent things will remain largely incomprehensible to humans, without a natural, intuitively understandable set of personalities. Establishing human-like trust between humans and things will be challenging at best (especially as intelligent things are susceptible to cyber intrusions that may compromise their 'perceptions' and decision making [3]). Things will also be relatively unsuccessful in negotiating with humans (but more successful when negotiating with other things [7]), and will find it hard to participate in defining a suitable role and task allocation in a mixed human-thing team.

# Control

Control involves adjustments to actions, which take into consideration changes in the situation and the progress or lack of progress that has been achieved.  It is this function of C2 that provides the agility necessary to be successful in fluid and dynamic situations.

In effect, control involves all other elements of the process in Figure 1, performed all over again and again – with agility -- as the execution of the solution faces inevitable breakdowns and unexpected stumbling blocks.  Humans face the possibility of information overload as they attempt to collect and absorb the relevant new information. Often, they feel disoriented when faced with unexpected and/or unfamiliar situations; then they take time to shift their mental models and may not "calculate / recalculate" quickly enough to accomplish the necessary coordinating actions. In some case, they are hobbled by either an actual or perceived need to obtain higher-level approvals for necessary changes.

This is where intelligent things may shine. They can and should do the control-related actions much faster. They are less likely to experience information overload, or psychological and cognitive barriers to 'overturning' past decisions to make a decision that is more appropriate for the situation at hand.  At the same time, precisely because of the in-humanly fast tempo of intelligent things, humans will find it difficult to understand and trust the recommendations or actions of things during an agile adaptation.

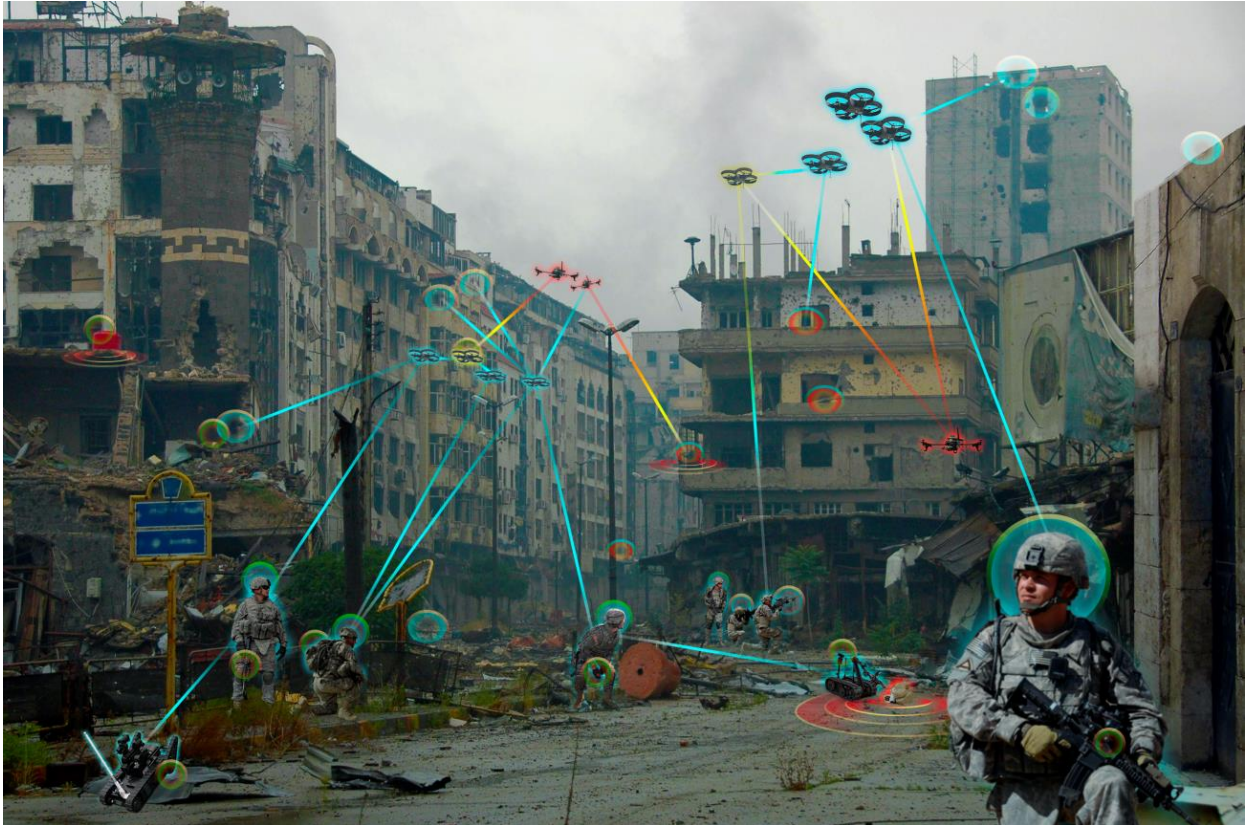

*Figure 2 This team of humans and intelligent things must make effective, distributed decisions in the dynamic, disorienting, dangerous environment. Graphics courtesy of: Tien Pham (concept) and Evan Jensen (art), U.S. Army Research Laboratory.*

## Sensemaking

In the case of organizations, making sense of the situation involves a number of iterative activities in the information, cognitive and social domains.  Sensemaking involves information seeking and analysis that is used to construct a story that represents our understanding of the situation, and frames our consideration of how to respond.  The components of Sensemaking include: obtaining relevant data, developing situation understanding, prediction, and decision making.  We will use these sensemaking activities to organize our look at how humans and other intelligent things can contribute to accomplishing these activities.

### Obtaining Relevant Data

To obtain relevant data, participants of a decision-making organization need to know what information is available, and how to access and process and filter it.  Creating high quality information to build situation awareness begins with shifting through large amount of data to select the most relevant (and timely) information, resolve conflicting information, and reject false information.

Humans are relatively slow in these processes, and can be easily overwhelmed when the quantity of information is extremely large. Facing the possibility of information overload, humans often focus on selected preferred sources and methods that they know well and trust. There is a tendency to collect, select and prefer the information that supports pre-existing biases. With information overload also comes increased error rate. Conflicting information could have come from multiple nodes within the organization, and reconciliation may become a matter of debates, personalities and office politics.

This is another area where intelligent things will shine. They bring high speed of information acquisition and processing, ability to deal with very large volumes of information, and consistent, rigorous and unbiased handling. Errors and omissions will be less likely, and conflicting data will be noticed. On the other hand, intelligent things rely largely on pre-existing models and algorithms (although a degree of learning will be expected), and are less likely to show creativity, to detect unconventional opportunity in data collection, or data relevance, or a clever deception hidden in the data.

## Situation Understanding

Understanding a situation involves far more that developing a description of the situation.  To support sensemaking our understanding needs to include perceptions of cause and effect, as well as temporal dynamics (how the situation is likely to unfold or change over time).   As humans, we say that we understand something when the result seems reasonable to us and we say that we do not understand it when the result is unexpected or (at least to us) without a logical explanation [2].

Understanding resides in the cognitive domain and, like everything in the minds of humans, is subjective, influenced by perceptual filters and biases. However, one's understanding may not be "correct," that is, it may not conform to objective reality. What does "understanding" mean in application to intelligent things?  Probably it involves the alignment of available data with models residing the thing's 'brain'; ensuring that observations of the environment can be explained by the available models of the environment. If so, things will be successful to the extent that a drastically novel model of reality is not required.

## Prediction

To understand something does not mean that one can predict a behavior or an event. Prediction requires more than understanding, thus even if one understands a phenomenon, one may not be able to predict, with anything that approaches a level of usefulness, the effect(s) of that phenomenon. Prediction requires actionable knowledge, specifically the values of the variables that determine (or influence) the outcome in question.  Operationally, the most that can be expected is to identify meaningfully different alternative futures and indicators that those alternatives are becoming more or less likely over time.

Humans have a reach, sophisticated based of experiential knowledge of what might happen in different situations. They have intuitive vision, the ability to project events forward. At the same time, each individual human member of a decision-making organization is limited in her/his knowledge. Alternative futures are debated, and a formal analysis is often distrusted or is too difficult to apply. As always, personalities and biases play important roles.

Things will be generally far better in applying formal or computational models for predictions [5], considering all pertinent details in an unbiased, exhaustive manner and drawing rigorous conclusions. However, they will have hard time explaining the rationale, the chain of reasoning, and articulating a compelling narrative that would illuminate their analysis of the situation [6].

### Decisionmaking

A number of diverse processes are comprised in decision-making: assess the pros and cons of the solutions, compare them, and articulate and defend the recommended solution. Too often, humans serve as enthusiastic carriers and champions of solutions looking for problems. As the saying goes, when you have a hammer, everything looks like a nail, that is, assessments of a solution are subconsciously biased towards the desired features or priorities of the problem, or the presumed preferences of a more senior decision-maker. For humans, disagreements about alternative solutions and their potential outcomes can devolve to a battle of personalities.

Intelligent things, on the other hand, will rigorously explore a broad space of potential solutions and match them to the requirements of the problem in an unbiased way, albeit within the scope of available machine knowledge, models and algorithms. If suitable analytical techniques exist, they will be used to characterize the advantages and disadvantages of the proposed solutions. However, as we already mentioned in another context earlier, an intelligent thing is likely to be challenged in explaining its chain of reasoning, to see "out of the box" solutions, or to recognize and manage the human emotional and political aspects in the competition of alternative solutions.

### Execution

Besides taking physical actions, or issuing orders for actions by subordinate entities, execution involves a complex process of coordinating details of the solution with other nodes within the decision making network: coordinating actions in time and space, managing interdependencies between and among actions, and resolving conflicts over resources. When adept in coordination (not always) humans achieve effective coordination by utilizing common understanding of goals and priorities (even if vaguely articulated), by following the established trusted links, and by negotiating a give-and-take deal when coordination if challenged by competition over priorities or resources. Humans are also capable of agile interpretations of orders if the right environment has been created.

These are skills that intelligent things have yet to develop. Things are not, at this point in their development, good at effective negotiation skills, especially when negotiating with humans; unable to understand unstated preferences and priorities of humans; or to build trust and mutual dependence relations.

## The C2 Challenge

This transformation of the 'workforce' promises to bolster our capabilities in a number of ways; perhaps most importantly by enhancing our agility. Yet, realizing these benefits will require developing appropriate ways to command this new organization form.   The fundamental challenge is one of C2 Design.   As more decisions and tasks migrate from humans to other intelligent entities, these entities need to be carefully integrated into our approaches to C2 in a manner that takes advantage of their unique qualities. Not to do so could lead to situations in which entities prevent each other from functioning as intended.

Best C2 practice calls for the allocation of decision rights to individuals to be based upon not only their competence to make the decision at hand, but also the ability to ensure that the decision maker can interact with others appropriately and has access to the information required. Adding other intelligent entities into the mix will require the explicit consideration of both the requirements of the decision task and the 'cognitive' attributes of these entities. The C2 Design challenge is to find the most appropriate allocation of decision tasks between and among humans and things for the mission and circumstances.  Although there is a desire to delegate responsibilities to the lowest level practical [9, 10], given that trust is necessary to achieve this in practice, how will this apply to intelligent things?

Commanders or managers of mixed human-thing organizations will face several challenges that the discussion above has highlighted.   Things are challenged in a number of areas and will need humans to provide these capabilities.  These include their ability to explain, build trust, bond, understand personal agendas, emotions, politics, and negotiate.   Things and people both to some extent have difficulty anticipating and coping with the unusual and unexpected and to think of out-of-the-box solutions.

Welcome aboard, intelligent things.  No matter what our respective shortcomings are, we will be stronger and more agile working together in decision-making organizations.

ALEXANDER KOTT is the Chief Scientist at the US Army Research Laboratory. Contact him at alexander.kott1.civ@mail.mil.

DAVID ALBERTS is a Senior Fellow at the Institute for Defense Analysis, where he explores issues related to command and control in a networked environment. Contact him at dalberts@ida.org.